\documentclass[letterpaper, 10 pt, conference]{ieeeconf}  

\IEEEoverridecommandlockouts                              

\title{\LARGE \bf
Systematic Stabilization of Constrained Piecewise Affine Systems
}

\author{Reza Lavaei$^{1}$, Leila Bridgeman$^{1}$
\thanks{$^{1}$Reza Lavaei and Leila Bridgeman are with the Department of Mechanical Engineering and Materials Science at Duke University, Durham NC, USA (email:  reza.lavaei@duke.edu; leila.bridgeman@duke.edu), corresponding author: Reza Lavaei}
}

\usepackage[utf8]{inputenc}
\usepackage{graphicx}
\usepackage{amsmath}
\usepackage{xfrac}
\DeclareMathOperator*{\argmin}{argmin}

\usepackage[table,xcdraw]{xcolor} 

\usepackage{gensymb}

\usepackage[nolist]{acronym}

\usepackage{braket}

\usepackage{physics}

\usepackage{amssymb,mathtools}
\usepackage{subcaption}

\usepackage[font=small,labelfont=bf]{caption}
\usepackage[font=small,labelfont=bf]{subcaption}

\usepackage{hyperref}

\usepackage{amsfonts}
\usepackage{nicefrac}

\usepackage{algorithm} 
\usepackage{algpseudocode}
\usepackage{cite}

\usepackage[final]{pdfpages}

\usepackage{pgfplots}
\pgfplotsset{compat=newest}
\pgfplotsset{plot coordinates/math parser=false} 

\usepackage{tabu}

\newtheorem{definition}{Definition}

\newtheorem{lemma}{Lemma}
\newtheorem{corollary}{Corollary}
\newtheorem{theorem}{Theorem}

\newtheorem{initialization}{Initialization}

\providecommand{\openbox}{\leavevmode
  \hbox to.77778em{%
  \hfil\vrule
  \vbox to.675em{\hrule width.6em\vfil\hrule}%
  \vrule\hfil}}
\makeatletter
\DeclareRobustCommand{\qed}{%
  \ifmmode
    \eqno \def\@badmath{$$}
    \let\eqno\relax \let\leqno\relax \let\veqno\relax
    \hbox{\openbox}%
  \else
    \leavevmode\unskip\penalty9999 \hbox{}\nobreak\hfill
    \quad\hbox{\openbox}%
  \fi
}
\makeatother
\newcommand{\mt}{m_{\mathcal{T}}}
\newcommand{\Et}{\mathbb{E}_\mathcal{T}}
\newcommand{\T}{\mathcal{T}}
\newcommand{\X}{\mathcal{X}}
\newcommand{\IntSet}{\mathbb{Z}}
\newcommand{\R}{\mathbb{R}}
\newcommand{\A}{\mathcal{A}}
\newcommand{\del}{\partial}
\newcommand{\C}{\mathcal{C}}

\newcommand{\Rg}{\mathcal{R}}

\begin{document}

\maketitle
\thispagestyle{empty}
\pagestyle{empty}

\begin{abstract}
This paper presents an efficient, offline method to simultaneously synthesize controllers and seek closed-loop Lyapunov functions for constrained piecewise affine systems on triangulated subsets of the admissible states. Triangulation refinements explore a rich class of controllers and Lyapunov functions. Since an explicit Lipschitz Lyapunov function is found, an invariant subset of the closed-loop region of attraction is obtained. Moreover, it is a control Lyapunov function, so minimum-norm controllers can be realized through online quadratic programming. It is formulated as a sequence of semi-definite programs. The method avoids computationally burdensome non-convex optimizations and a-priori design choices that are typical of similar existing methods.
\end{abstract}
\section{INTRODUCTION}
\Ac{PWA} state-space models can approximate a large class of nonlinear systems by partitioning the state-space into regions with distinct, affine dynamics \cite{sontag1981}. For instance, this can be done by linearizing a smooth nonlinear system around some operating points and selecting switching surfaces. Moreover, many hybrid systems have equivalent \ac{PWA} representations \cite{heemels2001,bemporad2004}. For most physical systems, respecting the state and input constraints must be ensured in control design. Consequently, systematic means to design stabilizing controllers would be broadly applicable and has garnered continual attention \cite{hassibi1998,rodrigues2003,mayne2003model,grieder2004low,lazar2006stabilizing,lazar2010,yordanov2011temporal,kaynama2012paper,samadi2014,stateDependent2017,marcucci2019,cabral2021stabilization}. This work builds upon the triangulation-based methods of \cite{gieslRevCPA2013,me} to create a novel, computationally efficient synthesis method for this important class of systems.

Stabilizing, piecewise linear state-feedback controllers and a single quadratic Lyapunov function can be sought for unconstrained PWA systems by solving \acp{LMI} using ellipsoidal approximations for the regions \cite{hassibi1998,mignone2000}. If unsuccessful, using exact region descriptions or searching in a richer class of functions is possible, however, the search often can no longer be formulated using \acp{LMI}. For instance, using exact polytopic descriptions trades-off against added conservatism due to the S-procedure \cite[Ch.\;3]{Liberzone} and getting a \ac{BMI} formulation \cite{hassibi1998}. Similarly, searching for \ac{PWA} state-feedback controllers \cite{rodrigues2004} and piecewise quadratic Lyapunov functions \cite[Ch\;6]{johansson1999} do not admit \ac{LMI} formulations, prompting strategies to tackle non-convexity that often require \textit{a-priori} choices or feasible initializations that cannot be made systematically. For instance, \cite{rodrigues2003} admits polytopic regions through the S-procedure and piecewise quadratic Lyapunov functions, but selects the closed-loop equilibria \textit{a-priori} and relies upon heuristics to seek a feasible initialization to pose a series of convex problems. Likewise, \cite{samadi2007} formulated the search for a quadratic Lyapunov function and \ac{PWA} state-feedback controllers by fixing the Lyapunov function's decay rate \textit{a-priori} and formulating a conservative convex optimization. Similar conservatism and restrictive convex-relaxations arise in \cite{lazar2010}. Even \cite{samadi2014}, which formulates \acp{LMI} and appears to circumvent initialization issues, demands a known \ac{CLF}.  

When respecting input and state constraints is critical, the problem deepens. The finite-time optimal controller has to account for possible switches and the set of initial states is generally non-convex, making it costly to find explicit offline solutions \cite{grieder2004low,Borrelli2005,lazar2006stabilizing,marcucci2019}. Guaranteeing recursive feasibility and Lyapunov stability in the receding-horizon implementation requires a terminal invariant set and a \ac{CLF} on that set \cite{di2014stabilizing,hariprasad2016,stateDependent2017}. If the closed-loop equilibrium is a shared point of some regions, finding such a \ac{CLF} is non-trivial and needs the methods of \cite{hassibi1998,rodrigues2003,samadi2007,lazar2010,samadi2014}, bringing once again the issues of the S-procedure, and initialization, and respecting constraints. 

Here, we propose an efficient controller synthesis method for state- and input-constrained \ac{PWA} systems that uses the \ac{CPA} controller and Lyapunov function selection techniques of \cite{me} to eschew \textit{a-priori} design choices and initialization heuristics. The system's \ac{PWA} nature removes the conservatism of \cite{me} in accounting for nonlinearities. The method circumvents the S-procedure entirely, removing the corresponding conservatism. Similar to \ac{EMPC}, online computation only involves evaluating a function. Moreover, the returned \ac{CLF} and the associated positive-invariant set of this method can be used as non-trivial terminal choices for \ac{EMPC}, removing a major challenge in synthesis. As pointed out in \cite{rantzer2000piecewise}, dividing the state-space into simplexes is a flexible way of control design as arbitrary functions' values can be interpolated by linear functions. We give an efficient implementation of this approach.

\section{Preliminaries}
\textbf{Notation.} The interior, boundary, and closure of $\Omega\in\R^n$ are denoted by $\Omega\degree$, $\del\Omega$, and $\bar{\Omega}$, respectively. The set of real-valued functions with $r$ times continuously differentiable partial derivatives over their domain is denoted by $\C^r$. The $i^{\textrm{th}}$ element of a vector $x$ is denoted by $x^{(i)}$. The preimage of a function $f$ with respect to a subset $\Omega$ of its codomain is defined by $f^{-1}(\Omega){=}\{x{\mid} f(x) {\in} \Omega \}$. The transpose and Euclidean norm of $x{\in}\R^n$ are denoted by $x^\intercal$ and $||x||$, respectively. The set of all compact subsets $\Omega{\subset}\R^n$ satisfying i) $\Omega\degree$ is connected and contains the origin, and ii) $\Omega=\overline{\Omega\degree}$, is denoted by $\mathfrak{R}^n$. The vector of ones in $\R^n$ is denoted by $1_n$. 

In this paper, the Lyapunov functions and controllers are defined on a triangulated subset of the state space. The required definitions are given next.

\begin{definition}[Affine independence{\cite{gieslRevCPA2013}}] \label{def:affDepVecs}
A collection of vectors $\{x_0,\ldots,x_n\}$ in $\R^n$ is called affinely independent if $x_1-x_0,\ldots,x_n-x_0$ are linearly independent. \qed
\end{definition}

\begin{definition}[$n$-simplex {\cite{gieslRevCPA2013}}] \label{def:simplex}
An $n$-simplex is the convex combination of $n+1$ affinely independent vectors in $\R^n$, denoted $\sigma{=}\textrm{co}(\{x_j\}_{j=0}^n)$, where $x_j$'s are called vertices. \qed
\end{definition}

\noindent In this paper, simplex always refers to $n$-simplex. By abuse of notation, $\T$ will refer to both a collection of simplexes and the set of points in all the simplexes of the collection. 

\begin{definition} [Triangulation {\cite{gieslRevCPA2013}}] \label{def:triangulation}
A set $\T\in\mathfrak{R}^n$ is called a triangulation if it is a finite collection of $\mt$ simplexes, denoted $\T=\{\sigma_i\}_{i=1}^{\mt}$, and the intersection of any of the two simplexes in $\T$ is either a face or the empty set. 

The following two conventions are used throughout this paper for triangulations and their simplexes. Let $\T=\{\sigma_i\}_{i=1}^n$. Further, let $\{x_{i,j}\}_{j=0}^n$ be $\sigma_i$'s vertices, making $\sigma_i=\textrm{co}(\{x_{i,j}\}_{j=0}^n)$. The choice of $x_{i,0}$ in $\sigma_i$ is arbitrary unless $0\in\sigma_i$, in which case $x_{i,0}=0$. The vertices of the triangulation $\T$ that are in $\Omega\subseteq\T$ is denoted by $\mathbb{E}_\Omega$. \qed
\end{definition}

\begin{definition} [Triangulable Set] \label{def:trinagulableSets}
A compact, connected subset of $\R^n$ that has no isolated points, and can be exactly covered by a finite number of simplexes. \qed
\end{definition}

\begin{definition} [Constraint Surfaces of a Triangulation]
Let $\T$ be the triangulation of a trinagulable set $\Omega\in\R^n$. The surface $\mathcal{H}\subset\T$ is called a constraint surface in $\T$ if it is exactly covered by the faces of some simplexes in $\T$. \qed
\end{definition}

\begin{lemma}[{\!\!\cite[Rem.\;9]{gieslRevCPA2013}}]\label{lem:nablaLinear}
Consider the triangulation $\T=\{\sigma_i\}_{i=1}^{\mt}$, where $\sigma_i=\textrm{co}(\{x_{i,j}\}_{j=0}^n)$, and a set $\mathbf{W}=\left\{ W_x \right\}_{ x\in \Et } \subset \R$. Let $\sigma_i=\textrm{co}(\{x_{i,j}\}_{j=0}^n)$, and $X_i\in\R^{n\times n}$ be a matrix that has $x_{i,j}-x_{i,0}$ as its $j$-th row, and $\bar{W}_i{\in}\R^n$ be a vector that has $W_{x_{i,j}}{-}W_{x_{i,0}}$ as its $j$-th element. The function $W(x)=x^\intercal_i X^{-1}_i \bar{W}_i+\omega_i$ is the unique, CPA interpolation of $\textbf{W}$ on $\T$, satisfying $W(x){=}W_x$, $\forall x{\in}\Et$. \qed
\end{lemma}

Note that since the elements of $\{x_{i,j}\}_{j=0}^n$ are affinely independent, $X_i$ in Lemma\;\ref{lem:nablaLinear} is invertible. The Dini derivative of a CPA $W$ at $x$ is defined as $D^+W(x) = \textrm{lim\,sup}_{h\rightarrow0^+}\sfrac{(W(x+hg(x))-W(x))}{h}$, which equals $\dot{W}(x)$ where $W\in\C^1$ \cite{gieslRevCPA2013}. Also, a continuous function $g(x)\in\R^n$ is piecewise in $\C^2$ on a triangulation $\T=\{\sigma_i\}_{i=1}^{\mt}$, denoted $g\in\C^2(\T)$, if it is in $\C^2$ on $\sigma_i$ for all $i\in\IntSet_1^{\mt}$ \cite[Def.\;5]{me}. The following theorem gives a general stabilization criteria for constrained systems using \ac{CPA} Lyapunov functions. 

\begin{theorem}[{\!\!\cite[Thm.\;3]{me}}] \label{thm:genOpt}
Consider the system
\begin{align} \label{eq:controlSystem}
    \dot{x} = g(x,u), \; x\in\X\in\mathfrak{R}^n, \; u\in\mathcal{U}\in\mathfrak{R}^m, \; g(0,0)=0.
\end{align}
Given a triangulation $\T =\{\sigma_i\}_{i=1}^{\mt}$, where $\T\subseteq\X$, suppose that a class of Lipschitz controllers $\mathcal{F}=\{u(\cdot,\boldsymbol{\lambda})\}$ parameterized by $\boldsymbol{\lambda}$ is chosen so that $u(0,\lambda)=0$, and $g_{\boldsymbol{\lambda}}(\cdot) \coloneqq g(\cdot,u(\cdot,\boldsymbol{\lambda}))$ is Lipschitz on $\T$, and both $u(\cdot,\boldsymbol{\lambda})$, $g_{\boldsymbol{\lambda}}(\cdot)\in\mathbb{C}^2(\T)$, and $u(\cdot,\boldsymbol{\lambda})\in\mathcal{U}$ for $\forall x\in\Et$ implies $u(\cdot,\boldsymbol{\lambda})\in\mathcal{U}$ for $\forall x\in\T$, and $\mathcal{F}$ has an admissible element. Consider the following nonlinear program.
\begin{subequations} \label{eq:genOpt} 
    \begin{alignat}{2}
        [\mathbf{V}^\ast,\; &\mathbf{L}^\ast,\; \boldsymbol{\lambda}^\ast,\; a^\ast,\; \mathbf{b}^\ast] = && \argmin_{\mathbf{V},\; \mathbf{L},\; \boldsymbol{\lambda},\; a,\; \mathbf{b}} \;\; \hat{J}(\mathbf{V}, \mathbf{L}, \boldsymbol{\lambda}, a, \mathbf{b})  \nonumber \\
        \textrm{s.t.}\;\;& V_0 = 0, \; a \geq 1,\; b_1 > 0, && \label{eq:genV0b1Constraint} \\
        & b_1||x||^a \leq V_x, &&\forall x\in\Et\backslash\{0\}, \label{eq:genMyVconstraint} \\
        & |{\nabla V}_i| \leq l_i, &&\forall i\in\IntSet_1^{\mt}, \label{eq:genNablaConstraint} \\
        & u(x_{i,j},\boldsymbol{\lambda})\in\mathcal{U}, &&\forall i\in\IntSet_1^{\mt}, \; \forall j\in\IntSet_0^n, \label{eq:genUconstraint} \\
        & D^+_{i,j}V \leq -b_2 V_{x_{i,j}}, \;\; &&\forall i\in\IntSet_1^{\mt}, \; \forall j\in\IntSet_0^n, \label{eq:genMyDv}
        \end{alignat}
\end{subequations}

\noindent where $D^+_{i,j}V=g_{\boldsymbol{\lambda}}(x_{i,j})^\intercal {\nabla V}_i + c_{i,j}\beta_i 1_n^\intercal l_i$, and $\mathbf{V}=\{V_x\}_{x\in\Et}\subset\R$ and $\mathbf{L}=\{l_i\}_{i=1}^{\mt}\subset\R^n$, and $\mathbf{b}=\{b_1,b_2\}\subset\R$, and $\hat{J}$ is a cost function, and for $u(\cdot,\boldsymbol{\lambda})$ satisfying \eqref{eq:genUconstraint},
\begin{flalign}\label{eq:betaAndc}
    &\beta_i \geq \max_{p,q,r\in\IntSet_1^n} \max_{\xi\in\sigma_i} \left| \left. \sfrac{\partial^2 g^{(p)}_{\boldsymbol{\lambda}}}{\partial x^{(q)}\partial x^{(r)}} \right|_{x=\xi} \right|, \textrm{ and} & \\
    &c_{i,j}{=} \frac{n}{2} ||x_{i,j} {-} x_{i,0}|| (\max_{k\in\IntSet_1^n} ||x_{i,k}{-}x_{i,0}|| {+} ||x_{i,j}{-}x_{i,0}||).& \nonumber
\end{flalign}

\noindent The optimization \eqref{eq:genOpt} is feasible. If $b_2^\ast > 0$ in \eqref{eq:genOpt}, then the CPA function $V^\ast:\T\rightarrow\R$ constructed from the elements of $\textbf{V}^\ast$ is a Lyapunov function of $\dot{x}=g_{\boldsymbol{\lambda}^\ast}(x)$. Let $\A = V^{\ast^{-1}}([0,r])\subseteq\T$ be in $\mathfrak{R}^n$ for some $r>0$. Then $x=0$ is locally exponentially stable for $\dot{x}=g_{\boldsymbol{\lambda}^\ast}(x)$ with $||x(t)||\leq\sqrt[^{a^\ast}]{r/b_1^\ast}e^{-(b_2^\ast/a^\ast)(t-t_0)}$ if $x(t_0)\in\A\degree$. \qed
\end{theorem}

\section{Main Results} \label{sc:mainResults}
By using \ac{CPA} functions for both the Lyapunov function and the controller, a method for stabilizing control-affine systems, $\dot{x}=f(x)+G(x)u$ with state and input constraints, was developed in \cite{me}. If $f(x)$ and $u$ are affine functions of $x$ on each simplex and $G(x)$ is constant, the $\beta_i$ term in \eqref{eq:betaAndc} vanishes since $g_\mathbf{\lambda}(x)$ is affine with respect to $x$. This observation leads to the following theorem that gives sufficient conditions for stabilization of \ac{PWA} systems.

\begin{theorem} \label{thm:genProg}
Consider the constrained control system
\begin{align} \label{eq:PWAsystem}
    \dot{x} = A_s x + B_s u + e_s, \;\;x\in\X\in\mathfrak{R}^n, \;\;u\in\mathcal{U}\in\mathfrak{R}^m,
\end{align}
where $\mathcal{U}=\Set{u\in\R^m \mid H u \leq h_c}$. Let $\Omega\in\mathfrak{R}^n\subseteq\X$, and $\{\Rg_s\}_{s = 1}^M$ be a partition of $\Omega$, where $A_s$, $B_s$, and $e_s$ are constant on each $\Rg_s$. In all $\Rg_s$'s containing the origin, $e_s=0$. Suppose that $\T$ is a triangulation of $\Omega$, comprised of triangulations of $\{\Rg_s\}_{s = 1}^L$. Let $u$ be CPA on $\T$, and $\mathbf{V}=\{V_x\}_{x\in\Et}\subset\R^n$, and $\mathbf{U}=\{u_x\}_{x\in\Et}\subset\R^m$, and $a\in\R$, and  $\mathbf{b}=\{b_1,b_2\}\subset\R$ be the unknowns. Consider

\begin{subequations} \label{eq:genProg} 
    \begin{alignat}{2}
        \mathbf{y}^\ast &= \argmin_{\mathbf{y}=  [\mathbf{V}, \mathbf{U}, a, \mathbf{b}]} \;\; J(\mathbf{y}) \nonumber \\
        \textrm{s.t.} \;\; & V_0 = 0, \; a \geq 1,\; b_1 > 0, && \label{eq:V0b1Constraint} \\
        & b_1||x||^a \leq V_x, && \forall x\in\Et\backslash\{0\}, \label{eq:myVconstraint} \\
        & u_0 = 0, \; H u_x \leq h_c, && \forall x \in \Et\backslash\{0\}, \label{eq:Uconstraint} \\
        & D^+_{i,j}V \leq -b_2 V_{x_{i,j}}, \quad && \forall i\in\IntSet_1^{\mt}, \; \forall j\in\IntSet_0^n,  \label{eq:Dv} 
        \end{alignat}
\end{subequations}

\noindent where $D^+_{i,j}V=(A_s x_{i,j} + e_s)^\intercal {\nabla V}_i + u_{x_{i,j}}^\intercal B_s^\intercal\nabla{V}_i$ and $J(\cdot)$ is a cost function. If $b_2^\ast > 0$ in \eqref{eq:genProg}, then the CPA function $V^\ast:\T\rightarrow\R$ constructed from the elements of $\textbf{V}^\ast$ is a Lyapunov function of $\dot{x}=A_s x + B_s u^\ast(x) + e_s$, where $u^\ast(\cdot)$ is the CPA function constructed from $\mathbf{U}^\ast$. Let $\A = V^{\ast^{-1}}([0,r])\subseteq\T$ be in $\mathfrak{R}^n$ for some $r>0$. Then $x=0$ is locally exponentially stable for the closed-loop system with $||x(t)||\leq\sqrt[^{a^\ast}]{r/b_1^\ast}e^{-(b_2^\ast/a^\ast)(t-t_0)}$ if $x(t_0)\in\A\degree$. \qed
\end{theorem}


\begin{proof}
We show that \eqref{eq:genProg} verifies \eqref{eq:genOpt}. By requiring $\T$ to be comprised of triangulations of $\{\Rg_s\}_{s = 1}^M$, the switching surfaces of \eqref{eq:PWAsystem} are constraint surfaces. So, on each simplex in Theorem\;\ref{thm:genOpt}, $f(x){=}A_s x {+} e_s$ and $G(x){=}B_s$, allowing $\beta_i{=}0$ in \eqref{eq:betaAndc} since $u$ is an affine function of $x$ on each simplex. Thus, \eqref{eq:Dv} verifies \eqref{eq:genMyDv} without needing \eqref{eq:genNablaConstraint}. Since $u$ is \ac{CPA}, \eqref{eq:Uconstraint} implies \eqref{eq:genUconstraint}. Lastly,  \eqref{eq:V0b1Constraint}--\eqref{eq:myVconstraint} are the same as \eqref{eq:genV0b1Constraint}--\eqref{eq:genMyVconstraint}. Thus the proof follows from that of Theorem\;\ref{thm:genOpt}.
\end{proof}

Even if $b_2^\ast \leq 0$ in \eqref{eq:genProg}, a connected subset of $\T$ that has $D_{i,j}^+V^\ast>0$ on its vertices might exist, using the following.

\begin{corollary}[{\!\!\cite[Cor.\;1]{me}}] \label{cor:mightFindAPosb2}
Suppose that $b_2^\ast\leq0$ in Theorem\;\ref{thm:genProg}. Let $\mathbb{I}= \{i\in\IntSet_1^{\mt} \mid D^+V^\ast_{x_{i,j}} < 0, \forall j\in\IntSet_0^n, x_{i,j}\neq0 \}$, and $\hat{\T}=\{\sigma_i\}_{i\in\mathbb{I}}$. Then, $V^\ast(x)$ satisfies $b_1^\ast||x||^{a^\ast}\leq V^\ast(x)$ and $D^+V^\ast(x)\leq \hat{b}_2^\ast V^\ast(x)$ for all $x\in\hat{\T}\degree$, where $\hat{b}_2^\ast \coloneqq \min \Set{-D^+_{i,j}V^\ast / V^\ast_{x_{i,j}} \mid i\in\mathbb{I}, j\in\IntSet_j^n, x_{i,j} \neq 0 }$. \qed
\end{corollary}

While Theorem\;\ref{thm:genProg} presents valid, sufficient conditions for closed-loop stability, bilinear terms $u_{x_{i,j}}^\intercal B_\lambda^\intercal \nabla{V}_i$ and $b_2V_{x_{i,j}}$ are present in \eqref{eq:Dv}, making \eqref{eq:genProg} non-convex. The following section presents more conservative criteria that are convex, enabling efficient iterative designs.

\section{Controller Design} \label{sc:ContDesign}
Algorithms that exploit Section\;\ref{sc:mainResults}'s criteria to design stabilizing controllers are proposed here. These entail two major steps: selecting adequately refined triangulations to admit stabilizing controllers while maintaining reasonable computational costs; and finding a \ac{CPA} controller and Lyapunov function over a given triangulation. The problem of finding the input and Lyapunov function is tackled first, and then the triangulation selection through refinement.

\subsection{Iterative Controller and Lyapunov Function Selection}
Closed-loop stability in \eqref{eq:genProg} is not ensured without a $b^\ast_2>0$. This section gives an iterative algorithm that seeks a controller and Lyapunov function satisfying the conditions of Theorem\;\ref{thm:genProg}. It iteratively increases $b_2$ until $b_2>0$ is found, ensuring closed-loop, exponential stability. Once $b_2>0$ is obtained, the algorithm fixes it, and optimizes other performance objectives. If either phase of improvement stagnates, triangulation refinement can be performed as discussed in Section\;\ref{sc:refinement}. However, before the iterative process can begin, an initial, feasible point of \eqref{eq:genProg} must be found, motivating the following initialization schemes.

\begin{initialization} \label{initialization:random}
Choose $a,b_1{>}0$, $V_x{=}b_1||x||^a$, $\forall x{\in}\Et$, $u_0{=}0$, and assign admissible $u_x$, $\forall x{\in}\Et\backslash\{0\}$, they can be random. Compute $\nabla{V}_i$ for all $i\in\IntSet_1^{\mt}$ as in Remark\;\ref{lem:nablaLinear}. Finally, find the largest $b_2$ satisfying \eqref{eq:Dv} in all simplexes. \qed
\end{initialization}

\begin{initialization} \label{initialization:LQR}
Design a LQR controller for one of the modes in \eqref{eq:PWAsystem} that has $e_\lambda=0$, and find the corresponding quadratic Lyapunov function, $x^\intercal\hat{P}x$. Sample $x^\intercal\hat{P}x$ at the vertices of $\T$ to find $\mathbf{V}$, and let $a{=}2$ and equate $b_1$ to the smallest eigenvalue of $\hat{P}$. Sample the LQR controller at the vertices of $\T$ to form $\mathbf{U}^{\textrm{LQR}}{=}\{u_x^{\textrm{LQR}}\}_{x\in\Et}$. Divide each element of $\mathbf{U}^{\textrm{LQR}}$ by a positive number so that the result, $\mathbf{U}{=}\{u_x\}_{x\in\Et}$, has admissible values for all vertices. Compute $\nabla{V}_i$ for all $i{\in}\IntSet_1^{\mt}$ as in Remark\;\ref{lem:nablaLinear} using the computed values of $V_x$ and $u_x$, respectively. Finally, find the largest $b_2$ satisfying \eqref{eq:Dv} in all simplexes. \qed
\end{initialization}

Each iteration in improving $b_2$ or other objectives is formulated by the following corollary.

\begin{corollary} \label{thm:semiProg2}
Suppose that $a>0$ is a fixed, known number in \eqref{eq:genProg}. Let $\underline{\mathbf{y}}=[\underline{\mathbf{V}}, \underline{\mathbf{U}}, a,  \underline{\mathbf{b}}]$ satisfy \eqref{eq:V0b1Constraint}--\eqref{eq:Dv}. Consider the following optimization.
\begin{subequations} \label{eq:SemiProg2} 
    \begin{alignat}{2}
        &\delta{\mathbf{y}^\ast} = \argmin_{\delta\mathbf{y} =  [\delta\mathbf{V}, \delta\mathbf{U}, 0, \delta\mathbf{b}]} && J(\underline{\mathbf{y}}+\delta\mathbf{y}) \nonumber \\
        &\textrm{s.t.} && \nonumber \\
        & \delta V_0 = 0, \;\; \underline{b}_1 + \delta b_1 > 0, && \label{eq:Semi2V1b0Constraint} \\
        & (\underline{b}_1+\delta b_1)||x||^a \leq \underline{V}_x +  \delta V_x, && \forall x\in\Et\backslash\{0\}, \label{eq:SemiVconstraint2} \\
        & \delta u_0 = 0, \; H (\underline{u}_x + \delta u_x) \leq h_c, && \forall x \in \Et, \label{eq:SemUconstraint2} \\
        & P_{i,j} \leq 0, && \forall i\in\IntSet_1^{\mt}, \; j\in\IntSet_1^n, \label{eq:2SemDv1}
        \end{alignat}
\end{subequations}

\noindent where $\delta{\nabla V}_i{=}X_i^{-1}\delta\bar{V}_i$, $\delta{\nabla u^{(s)}}_i{=}X_i^{-1}\delta\bar{u}_i$ as in Lemma\;\ref{lem:nablaLinear},
\begin{equation}
P_{i,j} = \begin{bmatrix} \phi_{i,j} & \ast & \ast & \ast & \ast \\
                          \delta{\nabla{V}}_i & -2I_n & \ast & \ast & \ast \\
                          B_s\delta u_{x_{i,j}} & 0 & -2I_n & \ast & \ast \\
                          \delta V_{x_{i,j}} & 0 & 0 & -2 & \ast \\
                          \delta b_2 & 0 & 0 & 0 & -2 \end{bmatrix},
\end{equation}

\begin{align}
    \phi_{i,j} = & ({\nabla \underline{V}}_i + \delta{\nabla V}_i)^\intercal(A_s x_{i,j} + B_s \underline{u}_{x_{i,j}}+ e_s) + \ldots \nonumber \\
    & {\nabla \underline{V}}_i^\intercal B_s \delta u_{x_{i,j}} + \underline{V}_{x_{i,j}}\delta b_2 + \ldots \nonumber \\
    & b_2 ( \underline{V}_{x_{i,j}}  + \delta V_{x_{i,j}}). \label{eq:phiHat} 
\end{align}

\noindent Then, $\underline{\mathbf{y}}+\delta\mathbf{y}^\ast$ is feasible for \eqref{eq:genProg}, and $J(\underline{\mathbf{y}}+\delta\mathbf{y}^\ast)\leq J(\underline{\mathbf{y}})$. \qed
\end{corollary}

\begin{proof}
To see \eqref{eq:SemiProg2}'s feasibility, observe that $\delta\mathbf{y}{=}0$ satisfies \eqref{eq:SemiProg2} since in this case, \eqref{eq:SemiProg2} is equivalent to \eqref{eq:genProg} with $\mathbf{y}{:=}\underline{\mathbf{y}}$. Substitution reveals that \eqref{eq:Semi2V1b0Constraint}--\eqref{eq:SemUconstraint2} imply \eqref{eq:V0b1Constraint}--\eqref{eq:Uconstraint} for $\mathbf{y}{=}\underline{\mathbf{y}}{+}\delta\mathbf{y}$. To see that \eqref{eq:2SemDv1} implies \eqref{eq:Dv}, note that $w^\intercal v {\leq} 1/2(w^\intercal w {+}v^\intercal v)$ for any two same-dimension vectors. Applying this fact with $(v,w){=}(\delta \nabla{V}_i, B_\lambda \delta u_{x_{i,j}})$ and $(v,w){=}(\delta V_{x_{i,j}}, \delta b_2)$ shows that by Schur Complement\cite[Ch\;2]{lmiBook}, \eqref{eq:2SemDv1} is implied. Finally, $J(\underline{\mathbf{y}} {+} \delta \mathbf{y}) {\leq} J(\mathbf{y})$ because otherwise $\delta\mathbf{y}{=}0$ would be a better, feasible solution.
\end{proof}

Given a triangulation and a linear or quadratic cost function $\hat{J}(\textbf{V},\textbf{U},b_1)$, a method of searching for a stabilizing \ac{CPA} controller is given in Algorithm\;\ref{alg:cpaControl}. It iteratively increases $b_2$ until it is positive. This can continue until a desired decay rate is ensured. Then, by fixing $b_2$'s value, $\hat{J}(\cdot)$ is iteratively minimized. If finding a positive $b_2$ is not successful, triangulation refinement, discussed later, is needed. 

\begin{algorithm}[h!]
	\caption{CPA control design on a fixed triangulation}
    \label{alg:cpaControl}
	\begin{algorithmic}[1]
	   \Require The \ac{PWA} system \eqref{eq:PWAsystem}, and a triangulation $\T\subseteq\X$ that has the switching surfaces as constraint surfaces, and a linear or quadratic $\hat{J}(\textbf{V},\textbf{U},b_1)$ 
	   \Ensure $u(x)$, and a positive-invariant set $\A$
	   \State $\underline{\mathbf{y}} \coloneqq $ a feasible point of \eqref{eq:genProg} (using Initialization\;\ref{initialization:random} or \ref{initialization:LQR})
	   \State $J \coloneqq -b_2$ \Comment{since $b_2$ is to be maximized}
	   \Repeat{}
	       \State Use Theorem\;\ref{thm:semiProg2}
	   \Until{$b_2 > 0$ is large enough OR $b_2$ is not changing} \label{line:term1}
	   \If{$b_2>0$ is found}
	       \State Fix $b_2$, and let $J\coloneqq\hat{J}(\cdot)$
	       \Repeat{}
	           \State Use Theorem\;\ref{thm:semiProg2}
	       \Until{$J$ is sufficiently small OR $J$ is not changing} \label{line:term2}
	       \State Return $u(x)$ and find a set $\A={V}^{-1}([0,r])$, $r>0$, \\ \hspace{0.4cm} where $\A\subseteq\T$ and $\A\in\mathfrak{R}^n$ \label{line:findingA}
	   \EndIf
	\end{algorithmic}
\end{algorithm}

Once Algorithm\;\ref{alg:cpaControl} returns, $\mathbf{y}$ can serve as the initial guess for \eqref{eq:genProg} with a desired cost function $J(\cdot)$ to improve performance offline. Moreover, since the corresponding Lyapunov function of the returned controller is also a Lipschitz CLF, a minimum-norm controller can be formulated as an online \ac{QP} \cite{Ames2016}. Suppose that $b_2^\ast>0$ is found by Algorithm\;\ref{alg:modification}, and $V^\ast$ is the corresponding CPA Lyapunov function. Let $\A$ be $\A{=}{V^\ast}^{-1}([0,r])$, $r{>}0$, where $\A{\subseteq}\T$ and $\A{\in}\mathfrak{R}^n$. Starting at any $x{\in}\A\degree$, the minimum-norm controller can be written as 
\begin{subequations} \label{eq:minNormLips} 
    \begin{flalign*} 
        &u^\ast(x) = \argmin_{u} \;\; u^\intercal\hat{H}(x)u + \hat{h}(x)^\intercal u & \nonumber \\
        &\textrm{s.t.\,} Hu\leq h_c, \, \nabla{V^\ast_i}^\intercal(A_s{+}B_s u{+}e_s){+} b_2^\ast V^\ast(x) {\leq 0}, \, \forall i{\in}\mathcal{I}, &
        \end{flalign*}
\end{subequations}

\noindent where $\mathcal{I} {=} \Set{i{\in}\IntSet_1^{\mt} | x{\in}\sigma_i}$, and $\hat{H}(x)$ is positive definite. The optimization is feasible for all $x{\in}\A$, because the corresponding CPA controller of $V^\ast$ is a feasible point for it.

\subsection{Triangulation Refinement} \label{sc:refinement}
If finding a $b_2>0$ in Theorem\;\ref{thm:genProg} or Algorithm\;\ref{alg:cpaControl} fails, the triangulation can be refined. This introduces more vertices and thus controller parameters, increasing the possibility of finding a stabilizing one. These refinements can be local by tracking the value of $D^+_{i,j}V$ on the simplexes in $\T$. An important assumption of Theorem\;\ref{thm:genProg} was that the switching surfaces of the PWA system are included in constraint surfaces of the triangulation. Let $\T$ be the triangulation of $\X$ in which $\mathcal{H}$, its constraint surfaces, include the switching surfaces of \eqref{eq:PWAsystem}, and let $\rho:\Omega\rightarrow\R_{>0}$, where $\Omega\subseteq\X$, be a function representing simplex sizes in a region of interest. Algorithm\;\ref{alg:modification} describes a simple way of refining triangulations.

\begin{algorithm}[h!]
	\caption{Control design with triangulation refinement}
    \label{alg:modification}
	\begin{algorithmic}[1]
	   \Require System \eqref{eq:PWAsystem}, cost function, simplex size function $\rho:\X\rightarrow\R_{>0}$, where $\hat{\Omega}\subseteq\Omega$, minimum simplex size $\rho_{\textrm{min}}$, constraint surfaces $\mathcal{H}$,  $0{<}\gamma{<}1$.
	   \Ensure $\mathbf{y} =  [\mathbf{V}, \mathbf{U}, a, \mathbf{b}]$
	   \Repeat{}
	        \State Generate $\T$, the $\X$'s triangulation respecting $\rho$ and $\mathcal{H}$, including the switching surfaces
	        \State Solve \eqref{eq:genProg} or use  Algorithm\;\ref{alg:cpaControl}
	        \If{desired objectives are met}
	            \State Return $\mathbf{y}$ \label{line:refFindingA}
	        \EndIf
	        \State $\rho := \gamma \rho$
	   \Until{$\rho_{\textrm{min}}$} is reached
	\end{algorithmic}
\end{algorithm}

\section{Numerical Simulation}
An example is adopted here from \cite[Sc\;5.4]{kaynama} with slight modifications, including additional constraints, to compare the introduced method, referred to as `CPA', with two other well-established ones, \ac{EMPC} and the \ac{PWA} method of \cite{rodrigues2003}. All the computations were carried out in MATLAB on a desktop computer with an AMD Ryzen\;5 CPU and 8\;GB DDR4 RAM. To solve \acp{SDP}, SeDuMi \cite{sedumi} with YALMIP \cite{yalmip} were used. The LQR cost for the initializations was $2x^\intercal x + u^\intercal u$. The toolbox Mesh2D \cite{mesh2d} was used for triangulation generation, where the maximum element size function was used for refinements.

Consider the PWA system \eqref{eq:PWAsystem} with $s\in\IntSet_1^3$, where $A_s=[0.1\;\;1.1;p_s\;\;-1]$, and $p_1=0.1$, $p_2=-0.9$, $p_3=-1.9$, and $B_s=[0\;\;1]^\intercal$, $\forall s\in\IntSet_1^3$, and $e_2=0$, and  $e_1=e_3=[0\;\;1]^\intercal$. The set $\X$, depicted in Fig.\;\ref{fig:case1ROA}, includes the polytopic regions $\{\Rg_s\}_{s=1}^3$, where $\Rg_1=\X\cap\{x\in\R^2\mid x^{(1)}\leq -1\}$, $\Rg_2=\X\cap\{x\in\R^2\mid -1\leq x^{(1)}\leq 1\}$, and $\Rg_3=\X\cap\{x\in\R^2\mid x^{(1)}\geq 1\}$. The input constraint is $|u|\leq u_{\textrm{max}}$. The problem is solved for the two cases, $u_{\textrm{max}}=1$ and $u_{\textrm{max}}=2$. The required offline time to synthesize stabilizing controllers, referred as synthesis time, denoted $t_{\textrm{syn}}^{\textrm{Ctrl}}$, and the settling times of closed-loop systems to $||x||\leq0.01$, denoted $t_{\textrm{settle}}^{\textrm{Ctrl}}$, and the ratio of the obtained \ac{ROA}'s area over $\X$'s area, denote $A^{\textrm{Ctrl}}_\X$, were compared using this paper's method and the two following ones.

\textbf{\ac{EMPC}}: The system was discretized using Euler's method with a $0.1$\;s sampling time. Since the origin is only in $\Rg_2$, the terminal set was selected as the maximal positive-invariant set in $\Rg_2$ associated with the terminal cost obtained from the solution of the Ricatti equation for the LQR cost $2x^\intercal x + u^\intercal u$, which was used as the running cost. Note that the terminal choices are not easy-to-find when the origin is shared between some regions. Thus, this example gives a significant advantage to EMPC's synthesis time. EMPC's horizon is denoted by $N$. For synthesis, MPT3 \cite{MPT3} was used.

\textbf{PWA \cite{rodrigues2003}}: This method searches for a quadratic Lyapunov function and a \ac{PWA} state-feedback, $u{=}K_sx{+}w_s$, in each $\Rg_s$, while maintaining continuity across the switching surfaces. We augmented it with $|K_sx_p{+}w_s|{\leq} u_{\textrm{max}}$, where $x_p$ denotes the vertices of $\Rg_s$, to enforce input constraints. Since \cite{rodrigues2003} did not address input constraints and would be computationally burdensome with more complex regions, this gives both additional functionality and an advantage to PWA's synthesis time. To solve the \acp{BMI}, \cite{rodrigues2003} alternates between fixing and seeking between finding Lyapunov functions versus controllers at each iteration. The function `fmincon' was used to find a feasible initialization for the controllers and the equilibria. Although finding suitable parameters to make \cite{rodrigues2003}'s method work involved some trial-and-error, they were not included in the synthesis time. The iterations increase a uniform decay rate for the Lyapunov functions.

\subsection{Case\;1: $u_{\textrm{max}}=1$}
\begin{figure}
\centering
    \begin{subfigure}[b]{\linewidth}            
            \includegraphics[width=\textwidth]{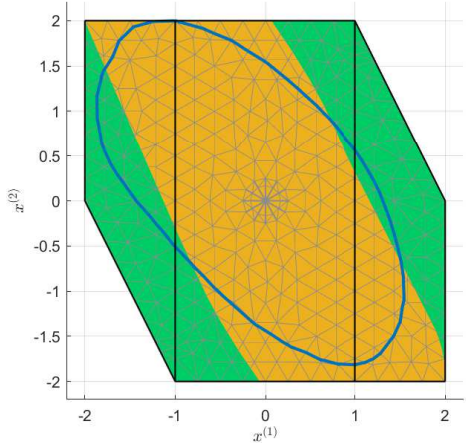}
            \caption{The set $\X$ (green), EMPC's ROA (yelllow), and the boundary of CPA's ROA (blue), and the CPA controller's triangulation (gray).}
            \label{fig:case1ROA}
    \end{subfigure}
    
    \begin{subfigure}[b]{\linewidth}
            \centering
            \includegraphics[width=0.8\textwidth]{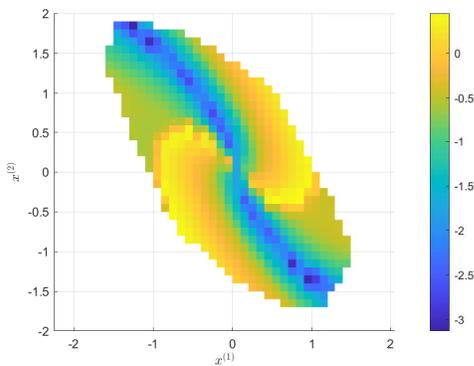}
            \caption{The value of $t^{\textrm{CPA}}_{\textrm{settle}}-t^{\textrm{EMPC}}_{\textrm{settle}}$ for same initial states.}
            \label{fig:case1Settle}
    \end{subfigure}
    \caption{Comparison of ROAs and settling times for Case\;1: $u_{\textrm{max}}=1$ between a CPA controller that has $b_2=1$ and EMPC with $N=5$.}\label{fig:case1Results}
\end{figure}

In this case, no PWA controller was found (this holds for $u_{\textrm{max}}{<}1.13$). Since both the CPA controller on the coarsest possible triangulation and EMPC with $N{=}1$ were able to find a stabilizing controller in about 2 seconds, we allowed them more time to achieve better performance. With $N{=}5$, EMPC found one in $73.2$\;s. A fine triangulation, depicted in Fig.\;\ref{fig:case1ROA} was generated for the CPA method. The values obtained by the CPA controller are compared to the EMPC in Table\;\ref{tab:my-table-U1}. The first stabilizing CPA was computed in about 7\;s and had comparable average settling time over the shared ROA, but its ROA's area was half of EMPC's. As $b_2$ increased, lower average settling times with respect to the EMPC were obtained, but although the ROA initially expanded, it eventually shrunk. Notably, during the same time needed to find the EMPC, the CPA formulates a controller that has comparable ROA area to that of EMPC with significant average settling time advantage. The results for this controller is visually compared to the EMPC in Fig\;\ref{fig:case1Results}. If slightly smaller ROA and advantage in settling time is preferred, CPA finds a controller in 0.6 of the time required for the EMPC, as in Table\;\ref{tab:my-table-U1}.

\begin{table}[]
\caption{Comparing the CPA controller on a fine triangulation to an EMPC that has $N=5$, $t_{\textrm{syn}}^{\textrm{EMPC}}=73.2\;s$, $A^{\textrm{EMPC}}_\X=0.64$}
\centering
\label{tab:my-table-U1}
\begin{tabular}{lcccc}
\hline
\rowcolor[HTML]{EFEFEF} 
\multicolumn{1}{|l|}{\cellcolor[HTML]{EFEFEF}$b_2$} &
  \multicolumn{1}{c|}{\cellcolor[HTML]{EFEFEF}0.07} &
  \multicolumn{1}{c|}{\cellcolor[HTML]{EFEFEF}0.88} &
  \multicolumn{1}{c|}{\cellcolor[HTML]{EFEFEF}1.00} &
  \multicolumn{1}{c|}{\cellcolor[HTML]{EFEFEF}1.17} \\ \hline
\multicolumn{1}{|l|}{$\nicefrac{ t_{\textrm{syn}}^{\textrm{CPA}} }{ t_{\textrm{syn}}^{\textrm{EMPC}} }$} &
  \multicolumn{1}{c|}{0.10} &
  \multicolumn{1}{c|}{0.63} &
  \multicolumn{1}{c|}{1.00} &
  \multicolumn{1}{c|}{1.69} \\ \hline
\multicolumn{1}{|l|}{$\nicefrac{ A^{\textrm{CPA}}_\X}{  A^{\textrm{EMPC}}_\X }$} &
  \multicolumn{1}{c|}{0.50} &
  \multicolumn{1}{c|}{0.83} &
  \multicolumn{1}{c|}{0.91} &
  \multicolumn{1}{c|}{0.51} \\ \hline
\multicolumn{1}{|l|}{$\nicefrac{ t_{\textrm{settle}}^{\textrm{av, CPA}} }{ t_{\textrm{settle}}^{\textrm{av, EMPC}} }^\ast$} &
  \multicolumn{1}{c|}{1.16} &
  \multicolumn{1}{c|}{0.87} &
  \multicolumn{1}{c|}{0.84} &
  \multicolumn{1}{c|}{0.81} \\ \hline
\multicolumn{5}{l}{$^\ast$ \small Average settling times over the shared ROA}
\end{tabular}
\end{table}


\subsection{Case\;2: $u_{\textrm{max}}=2$}
\begin{figure}
\centering
    \begin{subfigure}[b]{\linewidth}            
            \includegraphics[width=\textwidth]{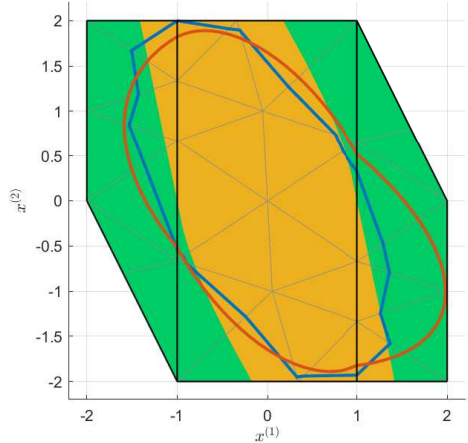}
            \caption{The set $\X$ (green), EMPC's ROA (yelllow), and the boundaries of CPA's ROA (blue), PWA's ROA (red), the switching surfaces (black), and the CPA controller's triangulation (gray).}
            \label{fig:case2ROA}
    \end{subfigure}
    
    \begin{subfigure}[b]{\linewidth}
            \centering
            \includegraphics[width=0.8\textwidth]{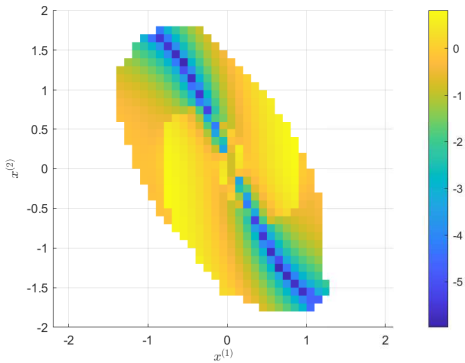}
            \caption{The value of $t^{\textrm{CPA}}_{\textrm{settle}}-t^{\textrm{PWA}}_{\textrm{settle}}$ for same initial states.}
            \label{fig:case2Settle}
    \end{subfigure}
    \caption{Comparison of ROAs of the three controllers, EMPC with $N{=}2$, CPA with $b_2{=}0.61$, and PWA, for Case\;2: $u_{\textrm{max}}{=}1$. The settling time is compared for the CPA and PWA controllers only.}\label{fig:case2Results}
\end{figure}

Here, PWA was also able to synthesize a controller as the input constraint was looser. Since it took only $4.9$\;s for PWA iterations to stagnate, we allowed the same synthesis time to CPA and EMPC. With $N{=}2$, EMPC synthesized a controller with $N{=}2$ in $2.8$\;s. Using the coarsest possible triangulation, depicted in Fig.\;\ref{fig:case2ROA}, the CPA controller achieved $b_2{=}0.6$ in $3.8$\;s. The values obtained by the CPA controller are compared to the PWA one in Table\;\ref{tab:my-table-U2}. The first CPA controller, obtained in only $0.6$\;s, had comparable average settling time over the shared ROA to the PWA's but its ROA's area is almost 25\% smaller. As $b_2$ increased, the CPA gained advantage in average settling time and expanded its ROA till stagnation at 6.5\;s. The obtained controller after 3.9\;s that had $b_2{=}0.61$ and 12\% settling time advantage over the PWA is visually compared to the PWA controller in Fig.\;\ref{fig:case2ROA}.

\begin{table}[]
\caption{Comparing the CPA controller on a coarse triangulation to a PWA one that has $t_{\textrm{syn}}^{\textrm{PWA}}=4.9\;s$ (stagnation), $A^{\textrm{PWA}}_\X=0.61$}
\centering
\label{tab:my-table-U2}
\begin{tabular}{lccccl}
\hline
\rowcolor[HTML]{EFEFEF} 
\multicolumn{1}{|l|}{\cellcolor[HTML]{EFEFEF}$b_2$} &
  \multicolumn{1}{c|}{\cellcolor[HTML]{EFEFEF}0.18} &
  \multicolumn{1}{c|}{\cellcolor[HTML]{EFEFEF}0.46} &
  \multicolumn{1}{c|}{\cellcolor[HTML]{EFEFEF}0.56} &
  \multicolumn{1}{c|}{\cellcolor[HTML]{EFEFEF}0.61} &
  \multicolumn{1}{l|}{\cellcolor[HTML]{EFEFEF}0.64} \\ \hline
\multicolumn{1}{|l|}{$\nicefrac{ t_{\textrm{syn}}^{\textrm{CPA}} }{ t_{\textrm{syn}}^{\textrm{PWA}} }$} &
  \multicolumn{1}{c|}{0.12} &
  \multicolumn{1}{c|}{0.25} &
  \multicolumn{1}{c|}{0.52} &
  \multicolumn{1}{c|}{0.80} &
  \multicolumn{1}{l|}{1.32} \\ \hline
\multicolumn{1}{|l|}{$\nicefrac{ A^{\textrm{CPA}}_\X}{  A^{\textrm{PWA}}_\X }$} &
  \multicolumn{1}{c|}{0.76} &
  \multicolumn{1}{c|}{0.78} &
  \multicolumn{1}{c|}{0.83} &
  \multicolumn{1}{c|}{0.84} &
  \multicolumn{1}{l|}{0.84} \\ \hline
\multicolumn{1}{|l|}{$\nicefrac{ t_{\textrm{settle}}^{\textrm{av, CPA}} }{ t_{\textrm{settle}}^{\textrm{av, PWA}} }^\ast$} &
  \multicolumn{1}{c|}{1.06} &
  \multicolumn{1}{c|}{0.97} &
  \multicolumn{1}{c|}{0.90} &
  \multicolumn{1}{c|}{0.88} &
  \multicolumn{1}{l|}{0.88} \\ \hline
\multicolumn{6}{l}{$^\ast$ \small Average settling times over the shared ROA}
\end{tabular}
\end{table}

\subsection{Discussion}
Although significant advantage was given to EMPC by including the origin in the interior of only one mode, and also to the PWA method by not accounting the time spent on trial-and-error, the CPA method was competitive to both and achieved significant improvements in terms of synthesis time and settling time. It always initializes feasibly, and removes \textit{a-priori} design choices. Moreover, triangulation refinement allows searching a rich class of Lyapunov functions and controllers as their defining vertices increase. This happens with no added conservatism and complexity, contrasting it with refinements that are also allowed in PWA method at the expense of added conservatism due to the S-prodecudre and more complexity in finding a feasible start. Both CPA and EMPC have rigorous ways of respecting input constraints. However, EMPC has a more sophisticated way of finding large \acp{ROA}. Using the CPA method to find terminal choices for EMPC when they are not trivial is appealing because of the discussed design convenience. Enlarging the \acp{ROA} using the CPA method will be considered in future.  

\section{Conclusion}
In this paper, a systematic, offline stabilization method for constrained \ac{PWA} systems was proposed that searches \ac{CPA} Lyapunov functions and controllers on triangulated subsets of the admissible states via iterative \acp{SDP}. The method returns an invariant subset of \ac{ROA} and explicit Lyapunov functions and controllers. Using the obtained Lipschitz CLF, an online \ac{QP}-based controller was also suggested. A comparison with two well-established methods was provided.

\bibliographystyle{unsrt}  
\bibliography{BibliographyShort.bib}

\begin{acronym}
\acro{SDP}{semi-definite program}
\acro{MPC}{model predictive control}
\acro{CLF}{control Lyapunov function}
\acro{CBF}{control barrier function}
\acro{CPA}{continuous piecewise affine}
\acro{QP}{quadratic programming}
\acro{DP}{dynamic programming}
\acro{ROA}{region of attraction}
\acrodefplural{ROA}{regions of attraction}
\acro{PWA}{piecewise affine}
\acro{LMI}{linear matrix inequality}
\acrodefplural{LMI}{linear matrix inequalities}
\acro{BMI}{bilinear matrix inequality}
\acrodefplural{BMI}{bilinear matrix inequalities}
\acro{EMPC}{explicit model predictive control}
\end{acronym}

\end{document}